\documentclass[runningheads]{llncs}
\usepackage[T1]{fontenc}
\usepackage{newtxtext}
\usepackage[varvw]{newtxmath}
\usepackage{graphicx}
\usepackage{amsmath}
\usepackage{booktabs}
\usepackage{orcidlink}
\usepackage{cite}
\usepackage{hyperref}
\hypersetup{hidelinks,pdftitle={Evolving dark energy after DESI: Nearby supernovae and an AdS future},pdfauthor={Ioannis D. Gialamas}}
\begin{document}
\title{Evolving dark energy after DESI: Nearby supernovae and an AdS future}
\titlerunning{Evolving dark energy after DESI: Nearby supernovae and an AdS future}
\author{Ioannis D. Gialamas\orcidlink{0000-0002-2957-5276}}
\authorrunning{I. D. Gialamas}
\institute{Laboratory of High Energy and Computational Physics, NICPB,\\
R\"avala pst. 10, 10143 Tallinn, Estonia\\
\email{ioannis.gialamas@kbfi.ee}}
\maketitle

\begingroup
\renewcommand{\thefootnote}{}
\footnotetext{\footnotesize Contribution to the proceedings of
\href{https://indico.cern.ch/event/1654737/}{HEP2026:
42nd Conference on Recent Developments in High Energy
Physics and Cosmology}, held in Athens.}
\endgroup

\begin{abstract}
We discuss the indications of evolving dark energy from fits to DESI baryon acoustic oscillation (BAO) measurements, compressed CMB data and the DES-SN5YR supernova sample. In fits using first-year DESI BAO data, excluding supernovae at $z<0.1$ substantially weakens the preference for evolving dark energy. The three-year DESI BAO measurements, however, favour a departure from $\Lambda$CDM when combined with CMB data alone.
Both truncated parametrizations and canonical scalar-field models fit the combined data well without requiring a phantom phase. To explore the implications for the future expansion of the Universe, we consider a simple scalar-field model with a Higgs-like potential that accommodates a recent decrease in the dark energy density. Given the adopted datasets and priors, the posterior probability of a negative potential minimum is $93.8\%$. Such a minimum would eventually lead to a reversal of the expansion and cosmological collapse.

\end{abstract}

\section{Introduction}
\label{sec:intro}

Recent baryon acoustic oscillation (BAO) measurements from the Dark Energy Spectroscopic Instrument (DESI) have renewed interest in whether the dark energy (DE) density evolves with time. In the standard $\Lambda$CDM model, dark energy takes the form of a cosmological constant with equation-of-state parameter $w=-1$. When combined with cosmic microwave background (CMB) and Type~Ia supernova (SN) data, both the first-year (DR1) and three-year (DR2) DESI measurements favour a time-dependent equation of state~\cite{DESI:2024mwx,DESI:2025zgx}. The strength of this preference depends on the choice of SN compilation, with DES-SN5YR, Union3 and Pantheon+ yielding different levels of significance~\cite{DES:2024jxu,Rubin:2023ovl,Scolnic:2021amr}.

The physical interpretation of these fits depends on how the evolution of $w$ is modelled. In the widely used Chevallier--Polarski--Linder (CPL) parametrization~\cite{Chevallier:2000qy,Linder:2002et}, the same two parameters determine the present value of $w$ and its evolution over the entire redshift range. A fit that captures a rapid change in $w$ at low redshift can therefore imply an earlier phantom phase with $w<-1$~\cite{Gialamas:2024lyw,Shlivko:2024llw,Nesseris:2025lke}. This raises the question of whether the data require such a phase or whether a canonical scalar field can account for the inferred evolution~\cite{Gialamas:2025fdv}.

In this contribution, we discuss this question using the fits of Refs.~\cite{Gialamas:2024lyw,Gialamas:2025fdv}. Hubble friction can keep a canonical scalar field nearly frozen close to a maximum of its potential for much of cosmic history, with $w\simeq-1$. As the Universe expands and the Hubble rate decreases, the field can begin to roll at late times. Its motion can thus account for a recent decrease in the dark energy density while maintaining $w\geq-1$ as long as the scalar energy density remains positive. A Higgs-like potential also allows us to follow the subsequent evolution of the Universe. If the potential minimum is negative, the Universe may eventually stop expanding and start contracting

We also consider the role of nearby supernovae in the preference for evolving dark energy~\cite{Gialamas:2024lyw,Efstathiou:2024xcq,Notari:2024zmi}. In Ref.~\cite{Gialamas:2024lyw}, we repeated the joint fit to DESI DR1 BAO, CMB and DES-SN5YR data after excluding supernovae at $z<0.1$. This cut substantially weakened the preference for evolving dark energy, indicating that nearby supernovae drive much of the preference in the DR1 analysis.

\section{Data and methodology}
\label{sec:data}

In a spatially flat Universe, the transverse comoving distance $D_M$, the Hubble distance $D_H$ and the luminosity distance $D_L$ are given by
\begin{equation}
    D_M(z) = c\int_0^z \frac{\mathrm{d}z'}{H(z')}\,, \qquad
    D_H(z) = \frac{c}{H(z)}\,, \qquad
    D_L(z) = (1+z)D_M(z)\,.
    \label{eq:distances}
\end{equation}
BAO measurements constrain distances in units of the sound horizon at the baryon drag epoch, $r_d$. Anisotropic BAO measurements constrain $D_M/r_d$ and $D_H/r_d$ separately, while isotropic measurements constrain $D_V/r_d$, where $D_V(z)=[zD_M^2(z)D_H(z)]^{1/3}$. SN data constrain the redshift dependence of $D_L(z)$ up to an overall magnitude normalisation. CMB data help distinguish the effects of DE evolution on these distances from those of the matter density and the overall distance scale.

For both DESI data releases, we use Gaussian likelihoods with the published covariance matrices. In the DR1 analysis~\cite{Gialamas:2024lyw}, we use 12 BAO measurements and treat $r_d h$ as a free parameter, where $h=H_0/(100\,\mathrm{km\,s^{-1}\,Mpc^{-1}})$. We include CMB information through the Planck constraint on the acoustic scale, $100\theta_{\rm MC}=1.04090\pm0.00031$~\cite{Planck:2018vyg}. In the DR2 analysis~\cite{Gialamas:2025fdv}, we use 13 BAO measurements~\cite{DESI:2025zgx} and a compressed CMB likelihood that constrains the acoustic angle and the physical baryon and matter densities~\cite{Bansal:2025ipo}. We assume standard early-Universe physics with one massive neutrino species of mass $0.06\,\mathrm{eV}$. Both analyses use a simplified treatment of the CMB likelihood.

For the fits that include SN data, both analyses use the DES-SN5YR compilation, which combines 1635 DES SNe with 194 nearby SNe at $z<0.1$~\cite{DES:2024jxu,Foley:2017zdq,Hicken:2009df,Hicken:2012zr}. We marginalise analytically over the overall magnitude normalisation~\cite{Goliath:2001af}. The nearby SNe provide a low-$z$ reference for the SN distance measurements. To assess their impact on the inferred DE evolution, we repeat the DR1 fits after excluding SNe at $z<0.1$~\cite{Gialamas:2024lyw}.

\section{Dynamical dark energy}
\label{sec:evolution}

For a separately conserved DE component with equation of state $w(a)$, the continuity equation gives
\begin{equation}
    \frac{\mathrm{d}\ln\rho_{\rm DE}}{\mathrm{d}\ln a} = -3\left[1+w(a)\right]\,, \qquad
    \rho_{\rm DE}(a) = \rho_{{\rm DE},0} e^{-3\int_1^a \left[1+w(a')\right]\frac{\mathrm{d}a'}{a'}}\,,
    \label{eq:continuity}
\end{equation}
thus, a positive DE density decreases as the Universe expands whenever $w>-1$. In the CPL parametrization, $w(a)=w_0+(1-a)w_a$, the parameter $w_a$ fixes both the slope of $w(a)$ and, for a given $w_0$, its high-redshift limit $w_0+w_a$. The fits favour a present-day equation of state with $w_0>-1$ and $w_a<0$, so that in the CPL parametrization $w$ decreases towards higher redshift and crosses into the phantom regime if its asymptotic value, $w_0+w_a$, falls below $-1$. To test whether the data require this phase, we consider models that allow $w$ to depart from $-1$ only at late times.

\begin{figure}[t]
    \centering
    \includegraphics[width=0.95\textwidth]{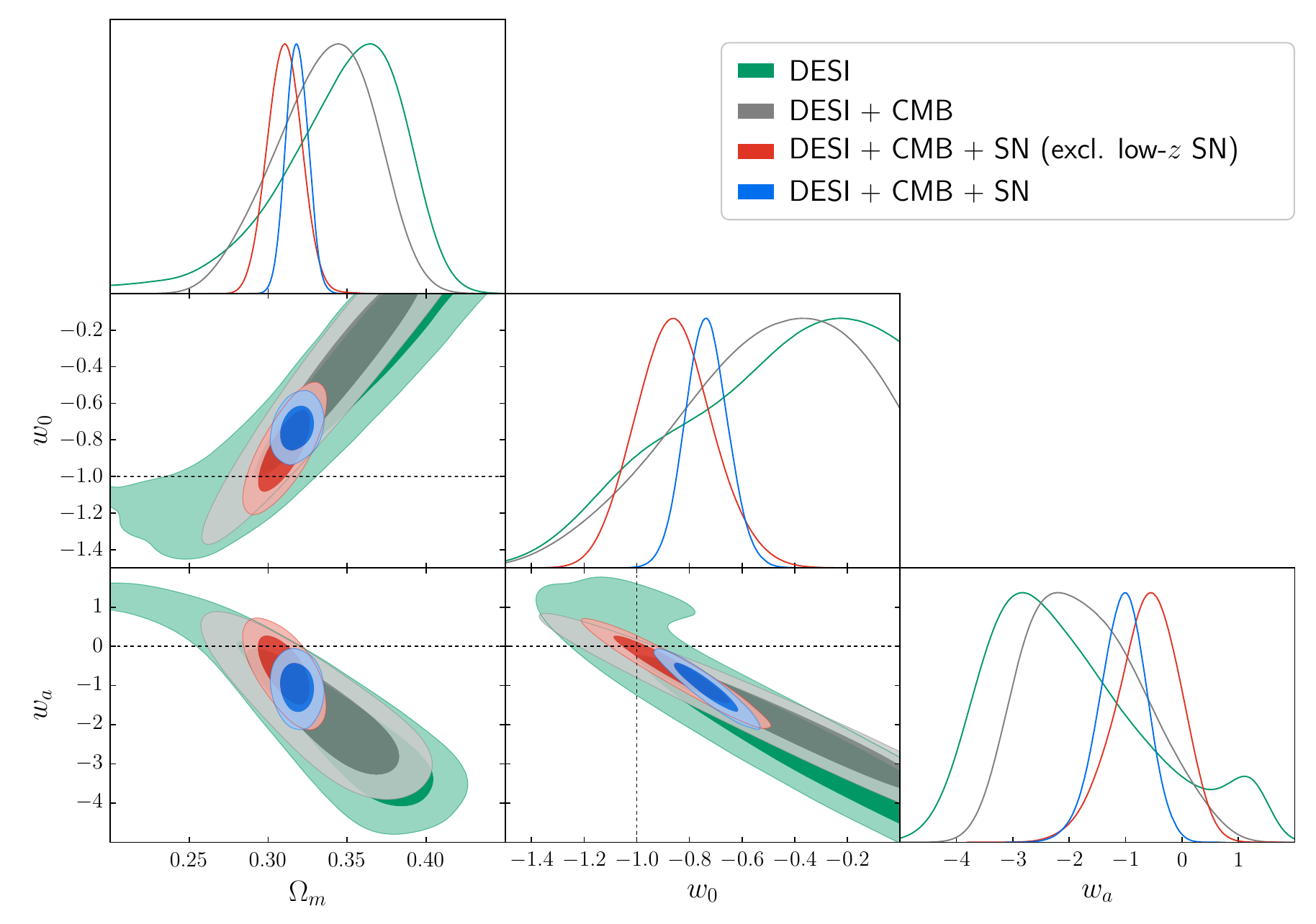}
    \Description{Constraints in the CPL parameter plane from DESI DR1, with CMB and SN data added successively. Excluding nearby SNe shifts the contours towards the cosmological-constant point.}
    \caption{Effect of excluding nearby SNe on the DR1 constraints in the $(w_0,w_a)$ plane. The contours enclose the $68\%$ and $95\%$ credible regions for the datasets indicated in the legend. The blue contours include the full SN sample, while the red contours exclude the 194 SNe at $z<0.1$. The dashed lines intersect at the $\Lambda$CDM point, $(w_0,w_a)=(-1,0)$. Reproduced from Fig.~1 of~\cite{Gialamas:2024lyw}.}
    \label{fig:lowz}
\end{figure}

In the DR1 analysis~\cite{Gialamas:2024lyw}, we compared CPL with a thawing quintessence model~\cite{Chiba:2009sj,Dutta:2008qn} and a step parametrization in which $w=w_0$ for $z<z_t$ and $w=-1$ for $z\geq z_t$. With the full SN sample, the minimum $\chi^2$ values are $1649.7$ for CPL, $1649.4$ for the step model and $1649.8$ for thawing quintessence, compared with $1662.6$ for $\Lambda$CDM. Despite their different forms of $w(z)$, all three best fits give a recent decrease of about $10\%$ in the DE density.

Excluding the 194 nearby SNe reduces the improvement over $\Lambda$CDM to $\Delta\chi^2=-0.9$ for CPL and about $-0.1$ for the other two models, with the shift in the CPL constraints shown in Fig.~\ref{fig:lowz}. The nearby sample, which extends to a comoving distance of roughly $300\,h^{-1}\,\mathrm{Mpc}$ at $z\simeq0.1$, thus accounts for much of the DR1 preference for evolving DE, although this cut alone cannot distinguish genuine DE evolution from effects that can influence the comparison with more distant SNe, such as peculiar velocities, calibration offsets or differences in the SN populations~\cite{Efstathiou:2024xcq,Notari:2024zmi}.

In the DR2 analysis~\cite{Gialamas:2025fdv}, we use the truncated CPL (tCPL) parametrization to set a floor $w_\infty$ that can vary independently of the late-time CPL slope,
\begin{equation}
    w_{\rm tCPL}(a) = \max \left\{w_0+(1-a)w_a,\,w_\infty\right\}\,.
    \label{eq:tcpl}
\end{equation}
Setting $w_\infty=-1$ gives tCPL($-1$), which approximates thawing quintessence, where Hubble friction keeps the field nearly frozen until it begins to roll at late times and $w$ rises above $-1$~\cite{Caldwell:2005tm,Tsujikawa:2013fta}. The Dutta--Scherrer--Chiba (DSC) approximation describes this evolution in terms of the present value of $w$ and the curvature of the potential, but becomes less accurate when the field rolls rapidly because it assumes a nearly constant DE density~\cite{Chiba:2009sj,Dutta:2008qn}.

\begin{table}[t]
    \centering
    \caption{Comparison of fits to DR2+CMB+DES-SN5YR~\cite{Gialamas:2025fdv}, with both differences evaluated relative to $\Lambda$CDM. The Akaike information criterion is $\mathrm{AIC}=\chi^2_{\min}+2k$, where $k$ is the number of fitted parameters.}
    \label{tab:fits}
    \begin{tabular}{l@{\hspace{1.6em}}r@{\hspace{1.3em}}r}
        \toprule
        Model & $\Delta\chi^2$ & $\Delta\mathrm{AIC}$ \\
        \midrule
        CPL               & $-17.2$ & $-13.2$ \\
        tCPL              & $-17.2$ & $-11.2$ \\
        tCPL($-1$)        & $-13.5$ & $-9.5$  \\
        DSC               & $-13.3$ & $-9.3$  \\
        Higgs-like scalar & $-13.5$ & $-7.5$  \\
        \bottomrule
    \end{tabular}
\end{table}

For DR2+CMB+DES-SN5YR, the models that allow a phantom phase lower $\chi^2$ by $17.2$ relative to $\Lambda$CDM, while models with $w\geq-1$ recover most of this improvement, with minimum $\chi^2$ values only $3.7$--$3.9$ higher (Table~\ref{tab:fits}). Although the best-fit tCPL value is $w_\infty\simeq-1.45$, the $95\%$ credible interval includes $-1$ under both priors, ranging from $[-1.55,-0.90]$ for a flat prior on $w_a$ to $[-1.81,-0.97]$ for a flat prior on $\log_{10}z_t$, where $z_t$ is the transition redshift. These fits favour DE evolution more strongly than a phantom phase, though a full Bayesian model comparison is still needed and some tCPL chains have limited effective sample sizes~\cite{Gialamas:2025fdv}.
\begin{figure}[t]
    \centering
    \includegraphics[width=0.82\textwidth]{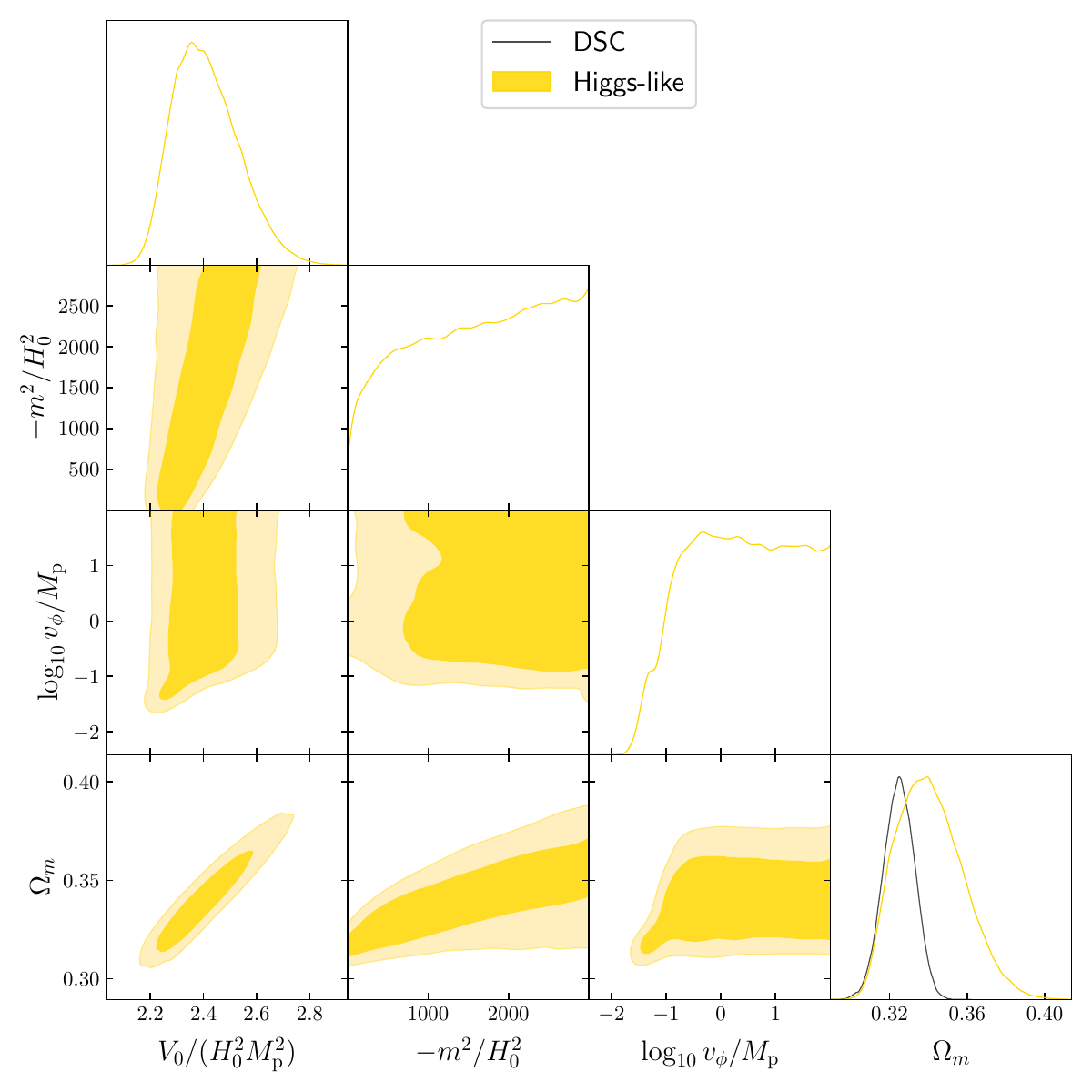}
    \Description{Posterior constraints on the height of the potential maximum, the mass parameter, the field value at the minimum and the matter density. The black marginal curve shows the DSC result for comparison.}
    \caption{Constraints on the Higgs-like model from DR2 BAO, compressed CMB data and DES-SN5YR, with yellow contours enclosing the $68\%$ and $95\%$ credible regions and the black curve showing the DSC result for $\Omega_m$. Reproduced from Fig.~4 of~\cite{Gialamas:2025fdv}.}
    \label{fig:scalar}
\end{figure}

\section{Higgs-like quintessence}
\label{sec:scalar}
\label{sec:fate}

We model DE as an ultralight canonical scalar field $\phi$ minimally coupled to gravity, with a Higgs-like potential
\begin{equation}
    V(\phi)=V_0+\frac{1}{2}m^2\phi^2+\frac{\lambda}{4}\phi^4\,,
    \qquad m^2<0\,, \quad \lambda>0\,.
    \label{eq:potential}
\end{equation}
The term ``Higgs-like'' refers to the shape of the potential, which has a local maximum at $\phi=0$ with height $V_0$ and two minima at $\phi=\pm v_\phi$, where $v_\phi=\sqrt{-m^2/\lambda}$. We leave $V_0$ free and use the quartic term to stabilise the potential, allowing us to follow the field as it rolls towards either minimum. Related quintessence potentials were studied in Ref.~\cite{Bhattacharya:2024kxp}.

The background equations are
\begin{align}
    \ddot\phi+3H\dot\phi+V_{,\phi} &= 0\,,
    & 3M_{\rm Pl}^2H^2 &= \rho_m+\rho_r+\rho_\phi\,,
    \label{eq:field}\\
    \rho_\phi &= \tfrac12\dot\phi^2+V(\phi)\,,
    & p_\phi &= \tfrac12\dot\phi^2-V(\phi)\,,
    \label{eq:stress}
\end{align}
where $M_{\rm Pl}$ is the reduced Planck mass, $\rho_m$ includes nonrelativistic neutrinos and $\rho_r$ is the radiation density. As the field rolls in an expanding Universe, its energy density decreases according to $\dot\rho_\phi=-3H\dot\phi^2$, while its equation of state satisfies $1+w_\phi=\dot\phi^2/\rho_\phi\geq0$ as long as $\rho_\phi>0$. A canonical field can thus account for a recent decrease in DE density without a phantom phase. If $V(\phi)$ becomes negative while $\rho_\phi$ remains positive, then $w_\phi>1$, and if $\rho_\phi$ subsequently crosses zero, the ratio $w_\phi=p_\phi/\rho_\phi$ diverges while the background equations remain regular.

We start the field at rest with $\dot\phi_i=0$ and $0<\phi_i<v_\phi$, determining $\phi_i$ by shooting for each set of parameters so that the solution reproduces the sampled value of $H_0$ today. Hubble friction initially keeps the field nearly frozen close to the maximum, and in the best-fit solution the DE density begins to decrease appreciably only at $z\lesssim0.1$, as in the nonphantom parametrizations. By today the field has crossed the minimum and begun to oscillate, with $w_{\phi,0}\simeq-0.14$~\cite{Gialamas:2025fdv}.

Figure~\ref{fig:scalar} shows that the potential parameters are weakly constrained, with best-fit values $V_0=2.27H_0^2M_{\rm Pl}^2$, $-m^2=2950H_0^2$ and $v_\phi\simeq0.039M_{\rm Pl}$, corresponding to a mass scale $\sqrt{-m^2}\simeq1.16\times10^{-31}h\,\mathrm{eV}$. The best-fit mass lies close to the upper prior bound $-m^2/H_0^2=3000$, and the solution requires a very small initial displacement, $\log_{10}(\phi_i/M_{\rm Pl})\simeq-46.1$, whose origin remains unexplained.

\begin{figure}[t]
    \centering
    \includegraphics[width=0.8\textwidth]{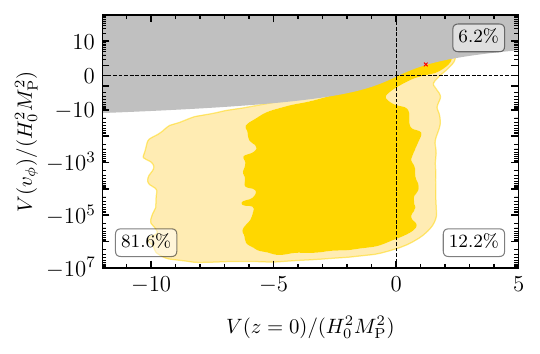}
    \Description{Joint posterior for the present scalar potential and its minimum. Most of the posterior lies at negative values of the minimum, while the best-fit point has both values positive.}
    \caption{Joint posterior for $V(\phi_{\rm today})$ and $V(v_\phi)$, with yellow contours enclosing the $68\%$ and $95\%$ credible regions and a red cross marking the best fit. The grey region violates $V(\phi_{\rm today})\geq V(v_\phi)$, and the percentages give the posterior probability in each allowed region. The vertical axis uses an inverse hyperbolic sine scale. Reproduced from Fig.~5 of~\cite{Gialamas:2025fdv}.}
    \label{fig:vacuum}
\end{figure}

The same potential that governs the recent motion of the field also determines its vacuum energy,
\begin{equation}
    V(v_\phi)=V_0+\frac{1}{4}m^2v_\phi^2
    =V_0-\frac{m^4}{4\lambda}\,.
    \label{eq:minimum}
\end{equation}
By fitting the recent evolution of $\rho_\phi$ within this model, we can infer the present potential $V(\phi_{\rm today})$ and its minimum $V(v_\phi)$, whose joint posterior is shown in Fig.~\ref{fig:vacuum}. The fit uses uniform priors in $V_0/(H_0^2M_{\rm Pl}^2)\in[0,6]$, $-m^2/H_0^2\in[0,3000]$ and $\log_{10}(v_\phi/M_{\rm Pl})\in[-5,2]$ (see~\cite{Gialamas:2025fdv}).

For the adopted data and priors, $93.8\%$ of the posterior has a negative minimum, with $81.6\%$ corresponding to a potential that is already negative today and $12.2\%$ to a potential that is still positive today. The best-fit point lies in the remaining $6.2\%$, where both values are positive, with $V(\phi_{\rm today})\simeq1.21H_0^2M_{\rm Pl}^2$ and $V(v_\phi)\simeq1.15H_0^2M_{\rm Pl}^2$.

A negative minimum corresponds to an anti-de Sitter (AdS) vacuum, but the total energy density today remains positive even if $V(\phi_{\rm today})<0$, because it also includes matter, radiation and the kinetic energy of the field. If the field continues to evolve in a potential with $V(v_\phi)<0$, the matter and radiation densities and the energy in scalar oscillations decrease as the Universe expands, until the total density reaches zero and Eq.~\eqref{eq:field} gives $H=0$. At that point, the second Friedmann equation,
\begin{equation}
    \dot H=-\frac{1}{2M_{\rm Pl}^2}
    \left(\rho_m+\frac{4}{3}\rho_r+\dot\phi^2\right)\,,
    \label{eq:hdot}
\end{equation}
gives $\dot H<0$, so expansion gives way to contraction. Once $H<0$, the scalar gains energy as it rolls and the Universe evolves towards collapse~\cite{Felder:2002jk}, although the posterior probability for $V(v_\phi)<0$ alone does not tell us when contraction would begin.

\section{Conclusions}

The DESI fits favour a recent decrease in the DE density, which canonical quintessence can describe without a phantom phase. In the DR1 analysis, excluding SNe at $z<0.1$ removes most of the improvement over $\Lambda$CDM. For DR2, BAO and CMB data alone already favour evolving DE, and canonical quintessence continues to fit the data well when DES-SN5YR is included.

For the Higgs-like potential, the DR2+CMB+DES-SN5YR fit gives a $93.8\%$ posterior probability for a negative minimum, conditional on the adopted initial conditions and priors. If the potential remains valid as the field evolves, such a minimum would eventually cause the Universe to stop expanding and start contracting.

Future DESI BAO measurements, together with further checks for systematic differences between nearby and distant SNe, should clarify whether the current preference for evolving DE persists.

\begin{credits}
\subsubsection{\ackname}
I am grateful to the colleagues with whom I carried out the work presented in Refs.~\cite{Gialamas:2024lyw,Gialamas:2025fdv}, on which this contribution is based. This work was supported by the Estonian Research Council grants PSG1132, TARISTU24-TK10 and TARISTU24-TK3, and the Centre of Excellence TK202 ``Fundamental Universe''. This article is based upon work from COST Action CA21136 (CosmoVerse), supported by COST (European Cooperation in Science and
Technology).

\subsubsection{\discintname}
The author has no competing interests relevant to this contribution.
\end{credits}

\bibliographystyle{splncs04-unsrt}
\bibliography{references}
\end{document}